\documentstyle[sprocl]{article}
\input{psfig}

\def\be{\begin{equation}}
\def\ee{\end{equation}}
\def\bea{\begin{eqnarray}}
\def\eea{\end{eqnarray}}

\def\lya{Ly$\alpha~$}

\def\tauh{\tau_{\rm HI}}

\begin{document}
\title{Characterization of Lyman Alpha Spectra
     and Predictions of Structure Formation Models: A Flux Statistics
Approach}
\author{ Rupert A.C. Croft, David H. Weinberg }
\address{Department of Astronomy, The Ohio State University,\\ Columbus,
 Ohio 43210, USA }
\author{ Lars Hernquist}
\address{Lick Observatory, University of California,\\
 Santa Cruz, CA 95064}
\author{ Neal Katz}
\address{Department of Physics and Astronomy, 
University of Massachusetts, \\ Amherst, MA, 98195}
\maketitle\abstracts{
In gravitational instability
 models, \lya  absorption arises from a continuous fluctuating medium,
so that spectra provide a non-linear one-dimensional ``map'' of the 
underlying density field.
 We characterise this continuous
absorption using statistical measures applied
 to the distribution of absorbed flux. 
We describe two simple members of a family of
statistics which we apply to simulated  spectra
in order to show their sensitivity as probes of
 cosmological parameters (H$_{0}$,
$\Omega$, the initial power spectrum of matter fluctuations) 
and the physical state of the IGM. We make use of 
 SPH simulation results to test the flux statistics, as well as presenting
 a preliminary application to Keck HIRES data.
}
\section{Introduction}

Hydrodynamical simulations are now providing us with the detailed predictions
of structure formation models for what should be seen in quasar absorption
spectra$^{1,2,3}$. If we accept  the same picture of formation of structure by
gravitional instability that is believed to be responsible for the 
observed galaxy distribution, then in the same sense that galaxy 
redshift surveys can provide important cosmological constraints, there is 
a wealth of information to be extracted from the \lya forest.

\begin{figure}
\vspace{4.4cm}
\includegraphics{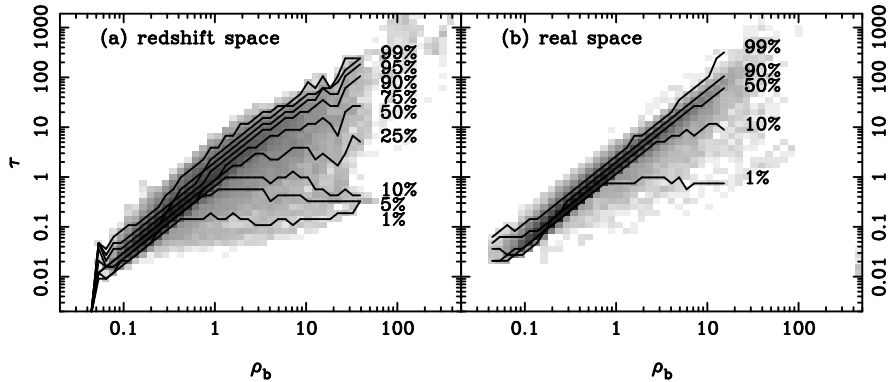}
\caption{ 
The joint distribution of optical depth and $\rho_b$ (in units of the mean
baryon density) for the SCDM model at $z=3.0$.
Panel (a) is in redshift space and in panel (b) peculiar velocities were
set to zero.
The logarithmic grey scale shows the fraction of pixels in
each  ($\rho_b$, $\tau$) bin.
Lines show the percentile distribution of $\tauh$ in each bin of $\rho_b$
\label{fig:scatter}}
\end{figure}

We analyse TreeSPH$^{4}$ hydrodynamic simulations of
three CDM-based cosmological models :
 SCDM and CCDM have $\Omega_{0}=1$, $h=0.5$ and 
$\sigma_{8}=0.7$, $\sigma_{8}=1.2$ respectively,
 OCDM has   $\Omega_{0}=0.4, \Omega_{\Lambda}=0, h=0.65, \sigma_{8}=0.75$.
 For all, $\Omega_{b}h^{2}=0.05$ 
and a photoionizing UV background is
included. Details of the simulations are given in [5].
In the models, \lya absorption arises from a continuous fluctuating medium,
the optical depth to absorption, $\tau$, being correlated with the
underlying density, which is shown in Figure 1. The empirically measured
relationship, $\tau \propto \rho_b^{1.6}$ can be predicted from  the 
interplay between  photoionization heating and adiabatic cooling$^{5,6}$.
  The scatter in
 1(a) is mainly due to the effects of peculiar velocities (set to 
zero in 1(b)).

\section{Flux statistics}
We apply two sets of statistics to the simulated spectra
as well as to a Keck spectrum of Q1422+231 (observations described in [7], 
 $\overline{z}$ of absorption = 3.2):  

(1) The fraction of each spectrum (FF, or filling factor)
above a given threshold in flux
decrement (D=$1-e^{-\tau}$). 
The dotted lines in Figure 2(a) show the results at two different redshifts
in the SCDM model. The  expansion of the Universe,
which results in a lower mean density of neutral hydrogen is the main 
factor driving the dramatic evolution in the shape of the curve and the
mean optical depth. All three models fit the
observations reasonably well (see also [8])
, once the strength of the UV background has
been adjusted to yield the correct mean amount of absorption, with CCDM
giving the best fit.

(2)  N, the mean number of times per unit length a spectrum crosses 
a given threshold in D. We plot N against against FF instead of D
(Figure 2(b)) so that the two panels give us entirely independent information.
When plotted in this fashion, N is member of
 a family of statistics such as the distribution
of distances between downcrossings
 and upcrossings (``size of absorbers'') which is
unaffected by variations in $\Omega_{b}$ and the UV background.
N is sensitive to 
the length scale and therefore H$_{0}$, as well as the power spectrum
of fluctuations and the temperature of the IGM (thermal broadening 
results in less crossings).

In Figure 2(b) we see that the three cosmological models have a smaller N
than  the Keck spectrum of Q1422+231, the difference being worse at high FF
(low D). Adding simulated Keck noise to the SCDM model does not resolve the
discrepancy. There is genuinely more structure in the QSO spectrum than
in low density regions of the simulations, . At the moment it is uncertain
whether  this discrepancy reflects a failure of these cosmological
models (e.g. they have the wrong power spectrum on these scales) or a failure
of the simulations to resolve the smallest scale features in the fluctuating
intergalactic medium.

\begin{figure}
\vspace{4.4cm}
\includegraphics{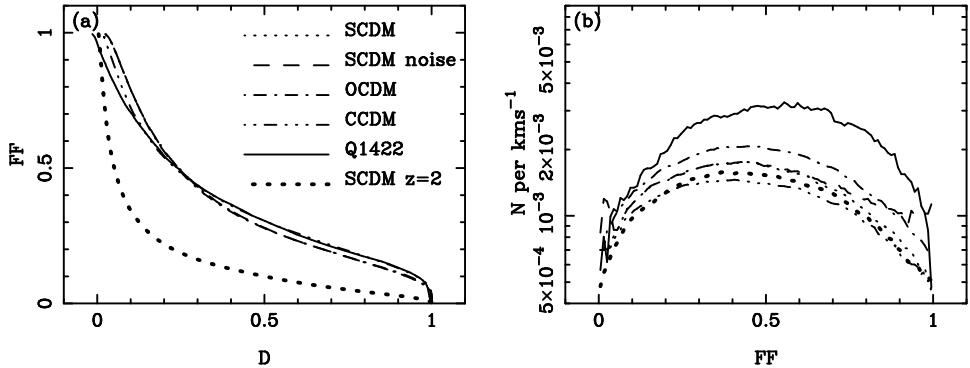}
\caption{ 
(a) Filling factor, FF vs flux decrement, D for 3 different models
(see text) at z=3, as well as SCDM with noise, SCDM at z=2 , and a Keck 
spectrum of Q1422+231. (b) Number of crossings, N, per kms$^{-1}$ vs FF.
 A 20 kms$^{-1}$ tophat filter 
was applied to all the spectra before calculating the statistics for 
both panels. 
\label{fig:stats}}
\end{figure}

\section*{Acknowledgments}
We thank Toni Songaila and Len Cowie for providing us with the
spectrum of Q1422+231
and Jordi Miralda-Escud\'{e} for helpul discussions.
The simulations were performed at the San Diego Supercomputer Center.

\section*{References}
1. J. Miralda-Escud\'{e} {\it et al}, ApJ {\bf 471},582(1996)\\
2. Y. Zhang {\it et al}, ApJ {\bf 453}, L57 (1995)\\
3. L. Hernquist {\it et al}, ApJ {\bf 457}, L51 (1996)\\
4. N. Katz {\it et al} ApJS {\bf 105}, 19 (1996)\\
5. R. Croft {\it et al}, ApJ {\it submitted}, astro-ph/9611053 (1996)\\
6. L. Hui and N. Gnedin, ApJ {\it in press}, astro-ph/9608157\\
7. A. Songaila, AJ {\it to be submitted} (1997)\\
8. M. Rauch {\it et al},  ApJ {\it submitted}, astro-ph/9612245 (1997)
\end{document}